# SEIR and Regression Model based COVID-19 outbreak predictions in India

Gaurav Pandey[a], Poonam Chaudhary[a], Rajan Gupta[b], Saibal Pal[c]

[a]Department of CSE & IT, The NorthCap University, India
[b]DeenDayalUpadhyaya College, University of Delhi, India
[c]Defence Research & Development Organization, India
[a]{Email: gaurav16csu120@ncuindia.edu}

**Abstract**
COVID-19 pandemic has become a major threat to the country. Till date, well tested medication or antidote is not available to cure this disease. According to WHO reports, COVID-19 is a severe acute respiratory syndrome which is transmitted through respiratory droplets and contact routes. Analysis of this disease requires major attention by the Government to take necessary steps in reducing the effect of this global pandemic. In this study, outbreak of this disease has been analysed for India till 30[th] March 2020 and predictions have been made for the number of cases for the next 2 weeks. SEIR model and Regression model have been used for predictions based on the data collected from John Hopkins University repository in the time period of 30[th] January 2020 to 30[th] March 2020. The performance of the models was evaluated using RMSLE and achieved 1.52 for SEIR model and 1.75 for the regression model. The RMSLE error rate between SEIR model and Regression model was found to be 2.01. Also, the value of $R_0$ which is the spread of the disease was calculated to be 2.02. Expected cases may rise between 5000-6000 in the next two weeks of time. This study will help the Government and doctors in preparing their plans for the next two weeks. Based on the predictions for short-term interval, these models can be tuned for forecasting in long-term intervals.
**Key-terms**: COVID-19, Corona Virus, India, Spread Exposed Infected Recovered model, Regression Model, Machine Learning, Predictions, Forecasting

## 1. Introduction

The COVID-19 (SARS-CoV-2) pandemic is a major global health threat. The Novel COVID-19 has been reported as the most detrimental respiratory virus since 1918 H1N1 influenza pandemic. According to the World Health Organization (WHO) COVID-19 situation report [1] as on March 27, 2020, a total of 509,164 confirmed cases and 23,335 deaths have been reported across the world. Global spread has been rapid, with 170 countries now having reported at least one case. Coronavirus disease 2019 (COVID-19) is an infectious disease caused by severe acute respiratory syndrome coronavirus 2 (SARS-CoV-2). Corona virus belongs to a family of viruses which is responsible for illness ranging from common cold to deadly diseases as Middle East Respiratory Syndrome (MERS) and Severe Acute Respiratory Syndrome (SARS) which were first discovered in China [2002] and Saudi Arabia [2012].
The 2019-novel Coronavirus or better known as COVID-19 was reported in Wuhan, China for the very first time on 31st December 2019. According to Jiang et al. [3] the fatality rate for this virus has been estimated to be 4.5% but for the age group 70-79 this has gone up to 8.0% while for those >80 it has been noted to be 14.8%. This has led to elderly persons above the age of 50 with underlying diseases like diabetes, Parkinson's disease and cardiovascular disease to be considered at the highest risk. Symptoms for this disease can take 2-14 days to appear and can range from fever, cough, shortness of breath to pneumonia, kidney failure and even death [1]. The transmission is person to person via respiratory droplets among close contact with the average number of people infected by a patient being 1.5 - 3.5 but the virus is not considered airborne [2].

There exist a large number of evidences where machine learning algorithms have proven to give efficient predictions in healthcare [4-6]. Nsoesie et al. [7] has provided a systematic review of approaches used to forecast the dynamics of influenza pandemic. They have reviewed research papers based on deterministic mass action models, regression models, prediction rules, Bayesian network, SEIR model, ARIMA forecasting model etc. Recent studies on COVID-19 include only exploratory analysis of the available limited data [8-10].

Effective and well-tested vaccine against CoVID-19 has not been invented and hence a key part in managing this pandemic is to decrease the epidemic peak, also known as flattening the epidemic curve. The role of data scientists and data mining researchers is to integrate the related data and technology to better understand the virus and its characteristics, which can help in taking right decisions and concrete plan of actions. It will lead to a bigger picture in taking aggressive measures in developing infrastructure, facilities, vaccines and restraining similar epidemics in future. The objectives of the current study are as follows.
1. Finding the rate of spread of the disease in India.
2. Developing a mathematical SEIR (Susceptible, Exposed, Infectious, Recovered) model to evaluate the spread of disease.
3. Prediction of COVID-19 outbreak using SEIR and Regression models.

After presenting background in Section I, Section II presents the methodology about the model used in this study. Section III covers analysis, experimental results and performance evaluation. Discussion has been provided in section IV followed by conclusion in section V.

## 2. Methodology

Time series data provided by John Hopkins University, USA has been used for the empirical result analysis [12]. The time period of data is from 30/01/2020 to 30/03/2020. The data includes confirmed cases, death cases and recovered cases of all countries. However, this paper focuses only on India's data for analysis and prediction of COVID-19 confirmed patients. The fact that India covers approximately 17.7% of the world's population and till date the effect of COVID-19 cases per million is less than 1, is the motivation behind this research. For analysis and prediction of number of COVID-19 patients in India, the following models have been used.

### 2.1 SEIR Model

Mathematical models can be designed to stimulate the effect of disease within many levels. These models can be used to evaluate disease from within the host model i.e. influence interaction within the cells of the host to metapopulation model i.e. how its spread in geographically separated populations. The most important part of this model is to calculate the $R_0$ value. The value of $R_0$ tells about the contagiousness of disease. It is the fundamental goal of epidemiologists studying a new case. In simple terms $R_0$ determines an average of what number of people can be affected by a single infected person over a course of time. If the value of $R_0 < 1$, this signifies the spread is expected to stop. If the value of $R_0 = 1$, this signifies the spread is stable or endemic. If the value of $R_0 >, 1$ this signifies the spread in increasing in the absence of intervention as shown in Figure 1. Equation (1) calculates the percentage of the population needed to be vaccinated to stabilize the spread of disease.

$$Population\ requires\ vaccine = 1 - \frac{Goal\ R0}{Current\ R0} \times Population \quad (1)$$

The R₀ value of COVID-19 for India that has been calculated based on earlier data is reported in the range 1.5 – 4 [11]. Assuming the goal of R₀ to be 0.5, the lower limit of the population required for the vaccine is 919.7 million and the upper limit would be 1.2071 billion. This remains a very vague figure so calculating the value of R₀ remains an important part.

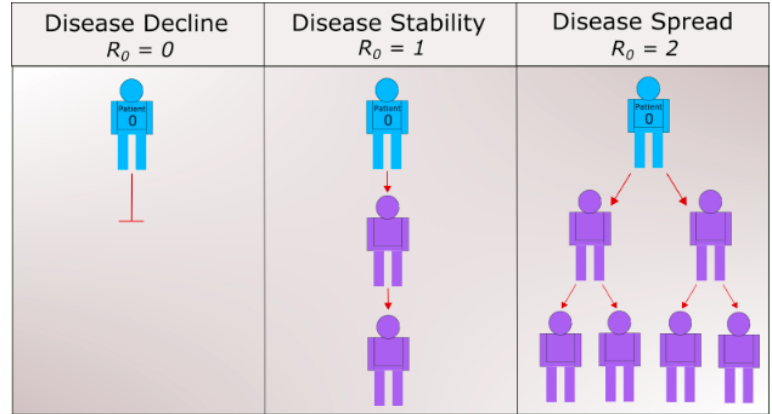

*Figure 1. Demonstration of spread of disease on the basis of $R_0$ value*

The SEIR model has mainly four components, viz. Susceptible (***S***), Exposed (***E***), Infected (***I***) and Recovered (***R***) as shown in Figure 2. ***S*** is the fraction of susceptible individuals (those able to contract the disease), ***E*** is the fraction of exposed individuals (those who have been infected but are not yet infectious), ***I*** is the fraction of infective individuals (those capable of transmitting the disease) and ***R*** is the fraction of recovered individuals (those who have become immune).

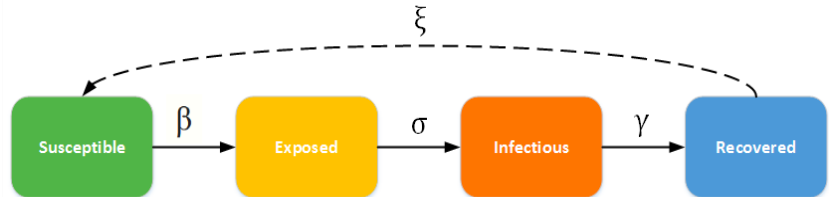

*Figure 2. Illustration of SEIR model and its four components*

In figure 2, ***β*** is the infectious rate which represents the probability of disease transmitting to susceptible person from an infectious person; ***σ*** is incubation rate which represents latent rate at which a person becomes infectious; ***γ*** is the recovery rate which is determined by $\frac{1}{D}$ (where D is duration of infection); and ***ξ*** is the rate at which recovered people become susceptible due to low immunity or other health related issues. The ordinary differential equations (ODEs) for all these four parameters are shown in equations 2 to 5.

$$\frac{dS}{dt} = -\frac{\beta SI}{N} + \xi R \qquad (2)$$

$$\frac{dE}{dt} = -\frac{\beta SI}{N} - \sigma E \qquad (3)$$

$$\frac{dI}{dt} = \sigma E - \gamma I \qquad (4)$$

$$\frac{dR}{dt} = \gamma I - \xi R \qquad (5)$$

Here, $N = S + E + I + R$ is the total the population. Now, we can calculate the value of $R_0$ using the formula in Equation 6. The value of $α, μ, β, γ$ can be calculated using the ordinary deferential equations from the 4 component of the SEIR model.

$$R_0 = \frac{\beta_0\, \alpha}{(\mu + \alpha)\,(\mu + \gamma)} \quad (6)$$

To describe the spread of COVID-19 using SEIR model, few consideration and assumptions were made due to limited availability of the data. They are enlisted as follows.

1. Number of births and deaths remain same
2. $1/\alpha$ is latent period of disease & $1/\gamma$ is infectious period
3. Recovered person was not sick again during the calculation period

Now, considering 70% of India's population to be approximately 966 million in susceptible class (**S**) and assuming only 1 person got infected in the initial part with average incubation period of 5.2, average infectious period of 2.9 and $R_0$ equal to 4, the SEIR model without intervention is shown in Figure 3 with the assumptions mentioned above. In Figure 3, we can examine that the number of susceptible population decreases by 80% in first 100 days as per the listed assumptions. The algorithm for SEIR model is shown as follows.

---

**Algorithm for SEIR Model**
**Input**: *Time Frame(T), Population(N), Number of infected cases(N_inf), Average incubation period(T_inc), Average Infectious period(T_inf), Reproduction number($R_0$), best decay function for $R_0(R_t)$*
**Output**: *S_out, E_out, I_out, R_out*
1. *S = (N − N_inf) / N*
2. *E = 0*
3. *I = N_inf / N*
4. *R = 0*
5. *For t < T do:*
6.     *S_out $= \frac{dS}{dt}$ (S, I, $R_t$, T_inf)*
7.     *E_out $= \frac{dE}{dt}$ (S, E, I, $R_t$, T_inf, T_inc)*
8.     *I_out $= \frac{dI}{dt}$ (I, E, T_inc, T_inf)*
9.     *R_out $= \frac{dR}{dt}$ (I, T_inf)*
10. *Return [S_out, E_out, I_out, R_out]*

---

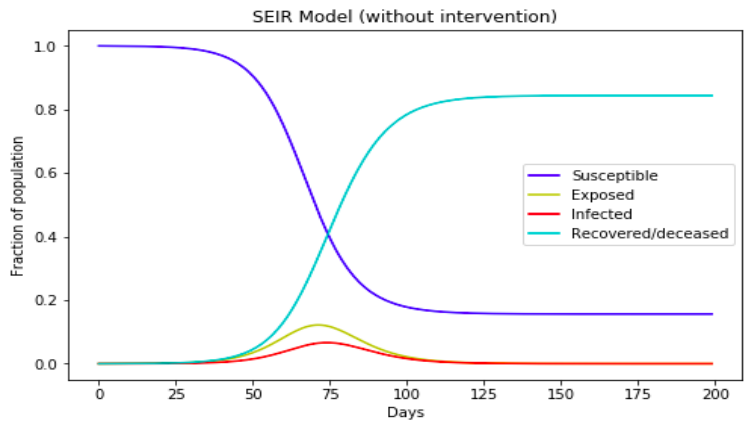

*Figure 3. SEIR Model using $R_0$ equal 4 and S to be 70% of India's population*

## 2.2 Regression Model

Regression models are statistical sets of processes which are used to estimate or predict the target or dependent variable on the basis of dependent variables. The regression model has many variants like linear regression, ridge regression, stepwise regression, polynomial regression etc. This study has used linear regression and polynomial regression for prediction of COVID-19 cases. Linear regression is a simple model which is used to finds the relation between a dependent and an independent variable. It uses the value of intercept and slope to predict the output variable. Equation 7 shows the relationship between a dependent and independent variable in a linear regression model. In Equation 7, $\boldsymbol{\beta_0}$ and $\boldsymbol{\beta_1}$ are two independent variables which represent intercept and slope respectively and $\epsilon$ is the error rate. This creates a straight line and is mostly used for predictive analysis. To make the linear regression algorithm more accurate we try to minimize the sum of residual square between the predicted and actual value.

$$Y = \beta_0 + \beta_1 x + \epsilon \tag{7}$$

Polynomial regression is a special type of regression which works on the curvilinear relationship between the dependent values and independent values. Equation 8 shows the relationship between a dependent and independent variable in polynomial regression. In Equation 8, $x$ is the independent variable and $\boldsymbol{\theta_0}$ is the bias also the intercept and $\boldsymbol{\theta_1, \theta_2}, \ldots, \boldsymbol{\theta_n}$ are the weight or partial coefficients assigned to the predictors and $\boldsymbol{n}$ is the degree of polynomial.

$$Y = \theta_0 + \theta_1 x + \theta_2 x^2 + \theta_3 x^3 + \theta_n x^n \tag{8}$$

The polynomial regression used in the study includes transformation of data into polynomials and applying linear regression to fit the parameter. A polynomial regression with degree equal to 1 is a linear regression. Choosing the value of a degree is a challenging task. If the degree of polynomial is less, it will not be able to fit the model properly and if the value of degree of polynomial is greater than actual, it will overfit the training data.

## 3. Results

In India, the first case of COVID-19 was reported on 30[th] January 2020. During the month of February, the number of cases reported was 3 and remained constant during the entire month. The major rise in the spread of disease started in the month of March'20. Figures 4 and 5 show the change in confirmed cases and death cases from 22[nd] Jan 2020 to 30[th] Mar 2020. Data from March shows a significant change in spread of the disease.

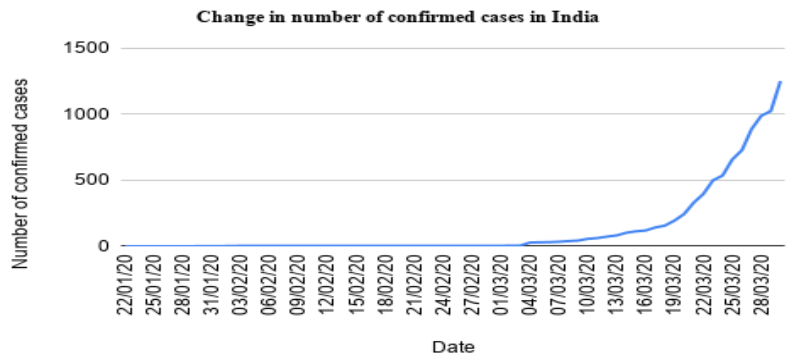

*Figure 4. Change in number of confirmed cases due to COVID-19 in India*

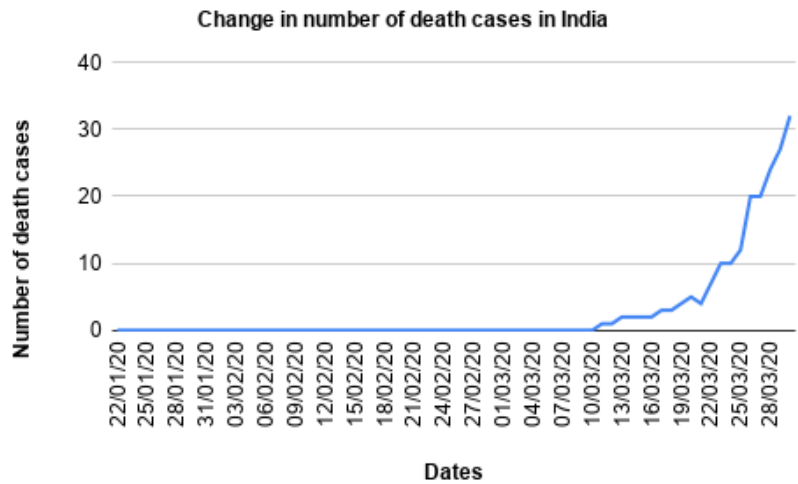

*Figure 5. Change in number of death cases due to COVID-19 in India*

In thecurrent analysis, we have used data till 25th March 2020 as our training data and data from 25th March 2020 - 30th March 2020 as the test/evaluation data. Before applying the prediction model we analysed the time series training data to check whether the model can fit the data or not. Figure 6 shows the result of confirmed cases in the $\log_{10}$ base versus last 15 days of training data. The reason of using last 15 days of training data is because it shows major growth in confirmed cases as shown in Figure 4.

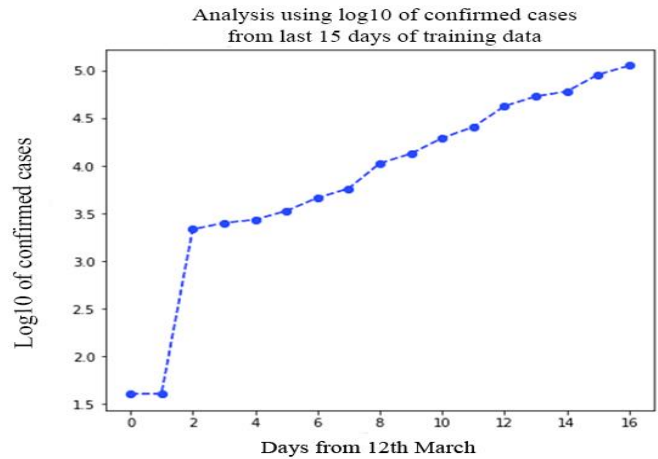

*Figure 6. Show the linear growth trend in India between 49 and 65 days*

Then we trained our data with SEIR model and result of line of fitting is shown in Figure 7. While applying SEIR, we took care of some interventions like quarantine and lockdown announced by the Government of India during this period so, we applied a decay function to lower the number of confirmed cases during prediction. We used hill decay in our model which is a half decay and its formula is given in Equation 9, where L is a description of the rate of decay, T is time and k is a shape parameter (no dimension). Half decay functions never reach zero and have half their original efficacies at time L.

$$hill\ decay\ function = 1/(1 + (T/L)^k) \qquad (9)$$

*Pandey et al.*

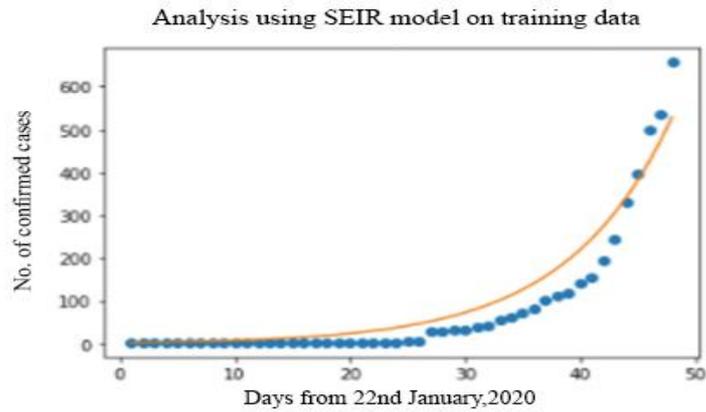

*Figure 7 Analysis using SEIR model, Blue dots are value from training data and yellow line is the model fitting line*

For India region, from the training data the value of beta was computed equal to 0.6206 and value of gamma equal to 0.5106. Now using Equation 2-6 we calculated the value of spread of disease or ***$R_0$ as equal to 2.02.***

Since we know COVID-19 is contagious and rate of transferring the disease from an infected person to susceptible person is 2.02 we need to predict the amount at which the disease can grow. Table 1 shows the prediction result using two models. In both the models, we have used 25 days training data as there was no significant trend in India before March 2020. To check the performance of the models used in this study we used root mean squared log error (RMSLE) and value RMSLE for the SEIR model was **1.52** and **1.75** for the regression model. The RMSLE error rate between SEIR model and Regression model was found to be **2.01**.

*Table 1. Prediction of number of COVID-19 patients in India for next two weeks using SEIR and regression model*

| Date | SEIR MODEL | Regression Model |
|---|---|---|
| 25/03/20 | 648 | 604 |
| 26/03/20 | 720 | 740 |
| 27/03/20 | 874 | 820 |
| 28/03/20 | 974 | 923 |
| 29/03/20 | 1081 | 1040 |
| 30/03/20 | 1199 | 1172 |
| 31/03/20 | 1333 | 1321 |
| 01/04/20 | 1485 | 1488 |
| 02/04/20 | 1654 | 1676 |
| 03/04/20 | 1837 | 1887 |
| 04/04/20 | 2038 | 2127 |
| 05/04/20 | 2264 | 2396 |
| 06/04/20 | 2520 | 2669 |
| 07/04/20 | 2807 | 3007 |
| 08/04/20 | 3122 | 3386 |
| 09/04/20 | 3465 | 3814 |
| 10/04/20 | 3847 | 4296 |
| 11/04/20 | 4279 | 4838 |
| 12/04/20 | 4763 | 5448 |
| 13/04/20 | 5300 | 6135 |

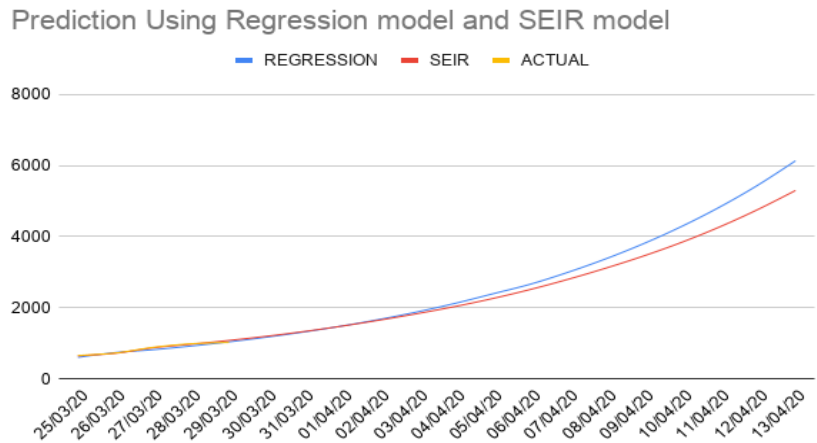

*Figure 8. Day-wise prediction using Regression [Blue], SEIR model [Red] and actual [Yellow] values. Actual values are recorded till 30/03/2020*

## 4. Discussion

The current trend shows that there will be a linear trend continued in the next few days as the control mechanisms taken by the Government of India are fairly strict and working well for the time being. Also, with linear trends, the patients getting recovered can be managed easily and death rate can be controlled as well. The findings of the current study may explode exponentially as shown by Gupta and Pal [13] if stringent control measures are not taken by the Government. Hospital provisions and medical facility enhancement work should be continued at a very rapid pace to prepare the country for exponential growth, if it occurs. However, with current interventions and preparations, the Government of India is looking forward to flatten the curve.

During the prediction of the confirmed cases as shown in Table 1, there were few challenges associated with the data. The data was not stationary and showed an exponential growth after 40 days from $22^{nd}$ Jan 2020 as shown in Figure 4. Overfitting remains a major problem with disease spread time series data. In this model, we have addressed the overfitting problem using decay based intervention. Another problem faced in this study was shortage of training data. Data for 25 days was used for training purpose and 5 days data for validation, based on which the number of confirmed cases for next 14 days was predicted. The training data is very less for any machine learning to train itself . Also, rapid changes in number of infected cases occurred in mid-March. The SEIR model shows an advantage as it does not grow exponentially with time but also uses some intervention methods with time. For intervention, a hill decay model was used. In the case of a regression model, different features can be used for decaying or intervention eg. the number of cases recovered etc., but growth of the regression line still remains a problem. For a regression model, we always need to train the model after some time with the change in trend in the data. The SEIR model also uses some assumptions like the number of people in susceptible class, for which we have taken 70% of the population to susceptible class.

In this study, we have only predicted the number of confirmed cases. To predict the number of death cases we faced many problems of data stationarity. Also, with limited data, the model was not able to predict the number of death cases properly. We have used only time series data for confirmed cases and death cases in this study. Using other data related to weather, geographic layout of the country, state-level population and governance parameters, the model prediction rate can be further improved.

## 5. Conclusion

In this study, two machine learning models SEIR and Regression were used to analyse and predict the change in spread of COVID-19 disease. We have analysed the data and found out that the number of cases per million in India is less than 0.5 till 30th March 2020. Then, with the help of the SEIR model the value of $R_0$ was computed to be 2.02. Also, we predicted the number of confirmed cases of COVID-19 for the next 14 days starting from 31st March 2020 – 13th April 2020. During performance evaluation, our model computed the value of RMSLE for the SEIR model to be **1.52** and **1.75** for the regression model. Also, the value of spread of disease of $R_0$ was found to be 2.02.

The result obtained from this study is taken from training data up to 30th March 2020. Further, looking at the trend, there is definitely going to be an increase in the number of cases. Doctors, health workers and people involved in providing essential services have to be protected in accordance with prescribed medical norms. Community spreading in the future due to carelessness of individuals as well as groups can exponentially increase the number of cases. The peak is yet to come, hence the Government has to be extra vigil and enforce strict measures. In addition, provision of medical facilities across the country has to be aggressively enhanced.

In future, an automated algorithm can be developed to fetch data in regular intervals and automatically predict the number of cases for weekly and bi-weekly data. In this way, Government and hospital facilities can also maintain a check on the supply and medical assistance / isolation required for new patients.